\documentclass[amsmath,amssymb,aps,letterpaper,prl,preprintnumbers,
twocolumn]{revtex4}

\usepackage[dvips]{graphicx}
\usepackage[dvips]{color}

\usepackage{hyperref}
\usepackage{amsfonts}

\usepackage{latexsym,amsmath,amssymb,graphics,stmaryrd}
\usepackage{array}

\usepackage{multirow}

\usepackage{pstricks}
\usepackage{pst-node}
\usepackage{pst-coil}

\usepackage{dsfont}
\usepackage{amsthm}
\usepackage{subfigure}
\usepackage{fancyhdr}

\makeatletter
\DeclareRobustCommand*{\bfseries}{%
\not@math@alphabet\bfseries\mathbf
\fontseries\bfdefault\selectfont
\boldmath
}
\makeatother

\newlength{\eqoff}

\newlength{\unit}
\psset{xunit=\unit,yunit=\unit,runit=\unit}
\newlength{\linew}
\psset{linewidth=\linew}

\newcommand{\col}{~,}
\newcommand{\pnt}{~.}
\newcommand{\e}{\operatorname{e}}
\DeclareMathOperator{\tr}{tr}

\newgray{ogray}{0.85}
\newgray{hatchgray}{1}
\newgray{sgray}{0.8}
\newgray{hiddengray}{0.9}

\newcommand{\olcolor}{ogray}
\newcommand{\olfillstyle}{crosshatch*}
\newcommand{\olhatchcolor}{red}
\newcommand{\orcolor}{ogray}
\newcommand{\orfillstyle}{crosshatch*}
\newcommand{\orhatchcolor}{red}
\newcommand{\sfillstyle}{solid}

\newlength{\rad}
\newlength{\roff}
\newlength{\ri}
\psset{xunit=\unit,yunit=\unit,runit=\unit}
\psset{linewidth=\linew}
\newlength{\dlinewidth}
\newlength{\doublesep}
\psset{doublesep=\doublesep}
\psset{linewidth=\linew}
\psset{dotscale=0.8}
\newlength{\auxlen}
\newlength{\linearc}
\newlength{\flinearc}
\newlength{\xa}
\newlength{\ya}
\newlength{\xb}
\newlength{\yb}
\newlength{\xc}
\newlength{\yc}
\newlength{\xd}
\newlength{\yd}
\newlength{\xe}
\newlength{\ye}

\newlength{\yf}

\newcommand{\ulinsert}[3][white]{%
\setlength{\xa}{#2\unit}
\addtolength{\xa}{0\unit}
\setlength{\xb}{#2\unit}
\addtolength{\xb}{1.5\unit}
\setlength{\xc}{#2\unit}
\addtolength{\xc}{3\unit}
\setlength{\ya}{#3\unit}
\addtolength{\ya}{0.5\dlinewidth}
\setlength{\yb}{#3\unit}
\addtolength{\yb}{-0.5\dlinewidth}
\setlength{\yc}{#3\unit}
\addtolength{\yc}{-1.5\unit}
\psset{doubleline=false}
\pscustom[fillstyle=\sfillstyle,fillcolor=#1,linecolor=#1,linewidth=0pt]{%
\psline[liftpen=1,linearc=\linearc](\xc,\yb)(\xb,\yb)(\xb,\yc)
\psline[liftpen=1](\xb,\ya)(\xc,\ya)}
\pscustom[fillstyle=\olfillstyle,fillcolor=\olcolor,hatchcolor=\olhatchcolor,
linecolor=\olcolor,linewidth=\linew]{%
\psline[linearc=\linearc](\xb,\yc)(\xb,\ya)
\psline[liftpen=1,linearc=2\linearc](\xb,\ya)(\xa,\ya)(\xa,\yc)}
\psline[linearc=\linearc](\xb,\yc)(\xb,\yb)(\xc,\yb)
\psline[linearc=2\linearc](\xa,\yc)(\xa,\ya)(\xb,\ya)
\psline(\xb,\ya)(\xc,\ya)
}
\newcommand{\dlinsert}[3][white]{%
\setlength{\xa}{#2\unit}
\addtolength{\xa}{0\unit}
\setlength{\xb}{#2\unit}
\addtolength{\xb}{1.5\unit}
\setlength{\xc}{#2\unit}
\addtolength{\xc}{3\unit}
\setlength{\ya}{#3\unit}
\addtolength{\ya}{-0.5\dlinewidth}
\setlength{\yb}{#3\unit}
\addtolength{\yb}{0.5\dlinewidth}
\setlength{\yc}{#3\unit}
\addtolength{\yc}{1.5\unit}
\psset{doubleline=false}
\pscustom[fillstyle=\sfillstyle,fillcolor=#1,linecolor=#1,linewidth=0pt]{%
\psline[linearc=\linearc](\xc,\yb)(\xb,\yb)(\xb,\yc)
\psline(\xb,\ya)(\xc,\ya)}
\pscustom[fillstyle=\olfillstyle,fillcolor=\olcolor,hatchcolor=\olhatchcolor,
linecolor=\olcolor,linewidth=\linew]{%
\psline[linearc=\linearc](\xb,\yc)(\xb,\ya)
\psline[liftpen=1,linearc=2\linearc](\xb,\ya)(\xa,\ya)(\xa,\yc)}
\psline[linearc=\linearc](\xb,\yc)(\xb,\yb)(\xc,\yb)
\psline[linearc=2\linearc](\xa,\yc)(\xa,\ya)(\xb,\ya)
\psline(\xb,\ya)(\xc,\ya)
}
\newcommand{\drinsert}[3][white]{%
\setlength{\xa}{#2\unit}
\addtolength{\xa}{0\unit}
\setlength{\xb}{#2\unit}
\addtolength{\xb}{-1.5\unit}
\setlength{\xc}{#2\unit}
\addtolength{\xc}{-3\unit}
\setlength{\ya}{#3\unit}
\addtolength{\ya}{-0.5\dlinewidth}
\setlength{\yb}{#3\unit}
\addtolength{\yb}{0.5\dlinewidth}
\setlength{\yc}{#3\unit}
\addtolength{\yc}{1.5\unit}
\psset{doubleline=false}
\pscustom[fillstyle=\sfillstyle,fillcolor=#1,linecolor=#1,linewidth=0pt]{%
\psline[linearc=\linearc](\xc,\yb)(\xb,\yb)(\xb,\yc)
\psline(\xb,\ya)(\xc,\ya)}
\pscustom[fillstyle=\orfillstyle,fillcolor=\orcolor,hatchcolor=\orhatchcolor,
linecolor=\orcolor,linewidth=\linew]{%
\psline[linearc=\linearc](\xb,\yc)(\xb,\ya)
\psline[liftpen=1,linearc=2\linearc](\xb,\ya)(\xa,\ya)(\xa,\yc)}
\psline[linearc=\linearc](\xb,\yc)(\xb,\yb)(\xc,\yb)
\psline[linearc=2\linearc](\xa,\yc)(\xa,\ya)(\xb,\ya)
\psline(\xb,\ya)(\xc,\ya)
}
\newcommand{\urinsert}[3][white]{%
\setlength{\xa}{#2\unit}
\addtolength{\xa}{0\unit}
\setlength{\xb}{#2\unit}
\addtolength{\xb}{-1.5\unit}
\setlength{\xc}{#2\unit}
\addtolength{\xc}{-3\unit}
\setlength{\ya}{#3\unit}
\addtolength{\ya}{0.5\dlinewidth}
\setlength{\yb}{#3\unit}
\addtolength{\yb}{-0.5\dlinewidth}
\setlength{\yc}{#3\unit}
\addtolength{\yc}{-1.5\unit}
\psset{doubleline=false}
\pscustom[fillstyle=\sfillstyle,fillcolor=#1,linecolor=#1,linewidth=0pt]{%
\psline[linearc=\linearc](\xc,\yb)(\xb,\yb)(\xb,\yc)
\psline[liftpen=1](\xb,\ya)(\xc,\ya)}
\pscustom[fillstyle=\orfillstyle,fillcolor=\orcolor,hatchcolor=\orhatchcolor,
linecolor=\orcolor,linewidth=\linew]{%
\psline[linearc=\linearc](\xb,\yc)(\xb,\ya)
\psline[liftpen=1,linearc=2\linearc](\xb,\ya)(\xa,\ya)(\xa,\yc)}
\psline[linearc=\linearc](\xb,\yc)(\xb,\yb)(\xc,\yb)
\psline[linearc=2\linearc](\xa,\yc)(\xa,\ya)(\xb,\ya)
\psline(\xb,\ya)(\xc,\ya)
}
\newcommand{\olvertex}[3][white]{%
\setlength{\xa}{#2\unit}
\addtolength{\xa}{0\unit}
\setlength{\xb}{#2\unit}
\addtolength{\xb}{1.5\unit}
\setlength{\xc}{#2\unit}
\addtolength{\xc}{3\unit}
\setlength{\ya}{#3\unit}
\addtolength{\ya}{1.5\unit}
\setlength{\yb}{#3\unit}
\addtolength{\yb}{0.5\dlinewidth}
\setlength{\yc}{#3\unit}
\addtolength{\yc}{-0.5\dlinewidth}
\setlength{\yd}{#3\unit}
\addtolength{\yd}{-1.5\unit}
\psset{doubleline=false}
\pscustom[fillstyle=\sfillstyle,fillcolor=#1,linecolor=#1,linewidth=0pt]{%
\psline[linearc=\linearc](\xc,\yb)(\xb,\yb)(\xb,\ya)
\psline[liftpen=1,linearc=\linearc](\xb,\yd)(\xb,\yc)(\xc,\yc)}
\pscustom[fillstyle=\olfillstyle,fillcolor=\olcolor,hatchcolor=\olhatchcolor,
linecolor=\olcolor,linewidth=0pt]{%
\psline[liftpen=0](\xa,\yd)(\xb,\yd)
\psline[liftpen=0](\xb,\ya)(\xa,\ya)
}
\psline[linecolor=\olcolor,linewidth=\linew](\xb,\ya)(\xb,\yd)
\psline[linearc=\linearc]{-C}(\xc,\yb)(\xb,\yb)(\xb,\ya)
\psline[liftpen=1,linearc=\linearc](\xb,\yd)(\xb,\yc)(\xc,\yc)
\psline{C-}(\xa,\ya)(\xa,\yd)
}
\newcommand{\orvertex}[3][white]{%
\setlength{\xa}{#2\unit}
\addtolength{\xa}{0\unit}
\setlength{\xb}{#2\unit}
\addtolength{\xb}{-1.5\unit}
\setlength{\xc}{#2\unit}
\addtolength{\xc}{-3\unit}
\setlength{\ya}{#3\unit}
\addtolength{\ya}{1.5\unit}
\setlength{\yb}{#3\unit}
\addtolength{\yb}{0.5\dlinewidth}
\setlength{\yc}{#3\unit}
\addtolength{\yc}{-0.5\dlinewidth}
\setlength{\yd}{#3\unit}
\addtolength{\yd}{-1.5\unit}
\psset{doubleline=false}
\pscustom[fillstyle=\sfillstyle,fillcolor=#1,linecolor=#1,linewidth=0pt]{%
\psline[linearc=\linearc](\xc,\yb)(\xb,\yb)(\xb,\ya)
\psline[liftpen=1,linearc=\linearc](\xb,\yd)(\xb,\yc)(\xc,\yc)}
\pscustom[fillstyle=\orfillstyle,fillcolor=\orcolor,hatchcolor=\orhatchcolor,
linecolor=\orcolor,linewidth=0pt]{%
\psline[liftpen=0](\xb,\ya)(\xa,\ya)
\psline[liftpen=0](\xa,\yd)(\xb,\yd)
\psline[liftpen=0](\xb,\ya)(\xa,\ya)
}
\psline[linecolor=\orcolor,linewidth=\linew](\xb,\ya)(\xb,\yd)
\psline[linearc=\linearc]{-C}(\xc,\yb)(\xb,\yb)(\xb,\ya)
\psline[liftpen=1,linearc=\linearc](\xb,\yd)(\xb,\yc)(\xc,\yc)
\psline{C-}(\xa,\ya)(\xa,\yd)
}

\newcommand{\fourvertex}[4][white]{%
\setlength{\xa}{0\unit}
\addtolength{\xa}{-1\unit}
\setlength{\xb}{0\unit}
\addtolength{\xb}{-0.5\dlinewidth}
\setlength{\xc}{0\unit}
\addtolength{\xc}{0.5\dlinewidth}
\setlength{\xd}{0\unit}
\addtolength{\xd}{1\unit}
\setlength{\ya}{0\unit}
\addtolength{\ya}{-1\unit}
\setlength{\yb}{0\unit}
\addtolength{\yb}{-0.5\dlinewidth}
\setlength{\yc}{0\unit}
\addtolength{\yc}{0.5\dlinewidth}
\setlength{\yd}{0\unit}
\addtolength{\yd}{1\unit}
\psset{doubleline=false}
\rput{0}(#2\unit,#3\unit){%
\pscustom[fillstyle=\sfillstyle,fillcolor=#1,linecolor=#1,linewidth=0pt]{%
\rotate{#4}
\psline[liftpen=1,linearc=\linearc](\xc,\ya)(\xc,\yb)(\xd,\yb)
\psline[liftpen=1,linearc=\linearc](\xd,\yc)(\xc,\yc)(\xc,\yd)
\psline[liftpen=1,linearc=\linearc](\xb,\yd)(\xb,\yc)(\xa,\yc)
\psline[liftpen=1,linearc=\linearc](\xa,\yb)(\xb,\yb)(\xb,\ya)}
\pscustom{%
\rotate{#4}
\psline[liftpen=1,linearc=\linearc](\xc,\ya)(\xc,\yb)(\xd,\yb)
\psline[liftpen=2,linearc=\linearc](\xd,\yc)(\xc,\yc)(\xc,\yd)
\psline[liftpen=2,linearc=\linearc](\xb,\yd)(\xb,\yc)(\xa,\yc)
\psline[liftpen=2,linearc=\linearc](\xa,\yb)(\xb,\yb)(\xb,\ya)
}
}
}

\newlength{\armlen}
\newcounter{nnodenum}
\newcounter{mnodenum}
\begin{document}


\preprint{HU-Mathematik-2016-02, \ HU-EP-16/06}

\title{On a CFT limit of planar $\gamma_i$-deformed $\mathcal{N}=4$ SYM theory
}

   \author{Christoph Sieg}%
    \email{csieg@physik.hu-berlin.de}
    \affiliation{%
Institute f\"ur Physik und Mathematik,
  Humboldt-Universit\"at zu Berlin,\\
  IRIS Geb\"aude, Zum Gro\ss{}en Windkanal 6, 12489 Berlin, Germany
}%
 \author{Matthias Wilhelm}%
  \email{matthias.wilhelm@nbi.ku.dk}
    \affiliation{%
Niels Bohr Institute, Copenhagen University,\\
Blegdamsvej 17, 2100 Copenhagen \O{}, Denmark
}%

\begin{abstract}
We show that an integrable four-dimensional non-unitary field theory
that was recently proposed as a certain limit of the $\gamma_i$-deformed $\mathcal{N}=4$ SYM theory is incomplete and not conformal -- not even in the 
planar limit. 
We complete this theory by double-trace couplings and find conformal one-loop 
fix-points when admitting respective complex coupling constants.
These couplings must not be neglected in the planar limit, as they can contribute to planar multi-point functions. Based on our results
for certain two-loop planar anomalous dimensions, we propose tests of integrability.
\end{abstract}

\maketitle

\section{Introduction}

In a recent paper \cite{Gurdogan:2015csr}, a certain limit is applied to the $\gamma_i$-deformation 
of $\mathcal{N}=4$ SYM theory \cite{Frolov:2005dj}, and the authors claim 
that the resulting non-unitary
theory is conformal in the planar limit where the number $N$ of 
colors is sent to infinity. 

In this letter, we point out that 
the Lagrangian given in \cite{Gurdogan:2015csr} is incomplete and does not define a 
CFT -- not even in the planar limit. 
We complete the theory by the missing couplings that are required for
renormalizability. They are quartic and have double-trace color structures and 
non-trivial $\beta$-functions \footnote{Quartic double-trace couplings emerged earlier in other 
contexts, e.g.\ in the effective potential 
of the worldvolume theory of interacting D3-branes \cite{Tseytlin:1999ii},
in non-supersymmetric orbifold theories \cite{Dymarsky:2005uh} where
they were related to tachyons in the dual string theory \cite{Dymarsky:2005nc}.}.
Although double-trace couplings are apparently subleading 
in the large-$N$ expansion, they must not be neglected in the planar limit
\cite{Sieg:2005kd}.
Planar multi-point correlation functions for infinitely many 
composite single-trace operators
depend on these couplings and are hence sensitive to the $\beta$-functions.  
We show this explicitly by determining to two-loop order 
the planar anomalous dimensions of several single-trace operators composed of 
two scalar fields.
Moreover, we give an example for a planar four-point correlation function of 
further operators that depends on one of the double-trace couplings.
Allowing for complex coupling constants in this model, whose single-trace
part is already non-unitary, we find (one-loop) fix-points.

\section{The proposed theory}

In \cite{Gurdogan:2015csr}, the authors propose to apply the following
limit to the $\gamma_i$-deformation
\begin{equation}\label{limit}
\gamma_i\to i\infty\col\quad \lambda\to0 \quad\text{with}\quad
\xi_i=\sqrt{\lambda}\e^{-\frac{i}{2}\gamma_i}=\text{const.} 
\end{equation}
and focus on the special case of only a single non-vanishing coupling constant 
$\xi=\xi_3$.
They claim that in the planar limit the Lagrangian of the resulting theory of two interacting complex 
scalars is given by equation (1) in their paper, 
which in our notation and conventions reads 
\begin{equation}
\begin{aligned}\label{L}
\mathcal{L}
&=\tr\Big(-\partial^\mu\bar\phi_1\partial_\mu\phi^1
-\partial^\mu\bar\phi_2\partial_\mu\phi^2
+\frac{\xi^2}{N}\bar\phi_1\bar\phi_{2}\phi^1\phi^{2}\Big)
\pnt
\end{aligned}
\end{equation}
This Lagrangian follows immediately when applying the limit \eqref{limit} to the single-trace part of the action of the $\gamma_i$-deformation as written e.g.\ in \cite{Fokken:2013aea}.

\section{Running double-trace couplings}

The Lagrangian \eqref{L} is incomplete since certain divergences in 
correlation functions of composite (single-trace) operators cannot be 
absorbed by renormalizing these operators -- not even in the planar limit.
These divergences have to be canceled by counter terms for quartic double-trace 
couplings.

At one-loop order, the counter terms are determined 
by contracting two copies of the quartic scalar single-trace vertex of \eqref{L}.
For fields that transform in the adjoint representation of the global $SU(N)$ group,
the only terms that have to be added to \eqref{L} read \footnote{For $U(N)$ fields, 
further couplings have to be considered \cite{Fokken:2013aea}.}
\begin{equation}
\begin{aligned}
\label{doubletracecouplings}
\mathcal{L}_{\text{dt}}&=
-\frac{\xi^2}{N^2}\bigg[\sum_{\substack{i\le j=1}}^2(Q^{ij}_{ij}+\delta Q^{ij}_{ij})\tr(\bar\phi_i\bar\phi_j)\tr(\phi^i\phi^j)\\
&\hphantom{{}={}-\frac{\xi^2}{N^2}\bigg[}
+(\tilde Q+\delta\tilde Q)\tr(\bar\phi_1\phi_2)\tr(\phi^1\bar\phi^2)
\bigg]
\col
\end{aligned}
\end{equation}
where $Q^{ii}_{ii}$, $Q^{12}_{12}$ and $\tilde Q$ with $i=1,2$ denote the respective
tree-level couplings. The counter terms 
$\delta Q^{ii}_{ii}$, $\delta Q^{12}_{12}$ and $\delta\tilde Q$ 
are easily determined in $D=4-2\varepsilon$ dimensions
and lead to the following one-loop $\beta$-functions
in the planar limit
\begin{equation}\label{betafunctions}
\begin{aligned}
\beta_{Q^{ii}_{ii}}&=2\varepsilon\delta Q^{ii}_{ii}
=(1+4(Q^{ii}_{ii})^2)\frac{\xi^2}{(4\pi)^2}
\col\\
\beta_{Q^{12}_{12}}&=2\varepsilon\delta Q^{12}_{12}=2(1-Q^{12}_{12})^2\frac{\xi^2}{(4\pi)^2}\col\\
\beta_{\tilde Q}&=2\varepsilon\delta\tilde Q=2(1-\tilde Q)^2\frac{\xi^2}{(4\pi)^2}
\pnt
\end{aligned}
\end{equation}

The $\beta$-functions \eqref{betafunctions} are non-vanishing for
$Q^{ii}_{ii}=Q^{12}_{12}=\tilde Q=0$, showing that the theory given by
the single-trace Lagrangian \eqref{L} alone is neither complete nor 
conformal.
The double-trace couplings 
\eqref{doubletracecouplings} cannot even be neglected in the planar limit: their counter terms
are required to renormalize certain planar Feynman diagrams. This is exemplified
below.

The $\beta$-functions for $Q^{12}_{12}$ and $\tilde Q$ 
have a (one-loop) fix-point at $Q^{12}_{12}=1$ and $\tilde Q=1$, respectively, 
such that their running can be avoided.
The $\beta$-functions for $Q^{ii}_{ii}$, however, do not have fix-points for real tree-level couplings $Q^{ii}_{ii}$.
In fact, the $\beta$-functions for $Q^{ii}_{ii}$ 
follow immediately when applying the limit
\eqref{limit} to our result \cite{Fokken:2013aea}, which we have 
obtained together with J.\ Fokken and which shows that
the $\gamma_i$-deformation is not a CFT. 
In contrast to the $\gamma_i$-deformation, a theory with the single-trace 
Lagrangian \eqref{L} is not unitary.
Hence,
let us be bold and even allow complex coupling constants.  
In this case, we can choose the imaginary values $Q^{ii}_{ii}=\pm\frac{i}{2}$ 
that are their one-loop fix-point values.

\section{Effect on the planar spectrum}

As we have already mentioned before, 
the double-trace couplings must not be neglected in the planar limit
since their counter terms are required to renormalize planar correlation functions. 
Together with J.\ Fokken, we
have already pointed this out in our calculation \cite{Fokken:2014soa} of the anomalous dimensions of the composite operators $\mathcal{O}^{ii}=\tr(\phi^i\phi^i)$
in the $\gamma_i$-deformation. 

Here, we determine to two-loop order 
the planar anomalous dimensions of all composite operators (states) built from 
two scalar fields
that receive contributions from the double-trace couplings and counter terms
\eqref{doubletracecouplings}. 
The affected states are $\mathcal{O}^{ii}=\tr(\phi^i\phi^i)$, $\mathcal{O}^{12}=\tr(\phi^1\phi^2)$, $\tilde{\mathcal{O}}=\tr(\bar\phi_1\phi^2)$ 
and their complex conjugates.
The results for their anomalous dimensions read
\begin{equation}
\begin{aligned}\label{gammas}
\gamma^\varrho_{\mathcal{O}^{ii}}
&=4Q^{ii}_{ii}\frac{\xi^2}{(4\pi)^2}-2\frac{\xi^4}{(4\pi)^4}-2\varrho\beta_{Q^{ii}_{ii}}\frac{\xi^2}{(4\pi)^2}\col\\
\gamma^\varrho_{\mathcal{O}^{12}}&=2(Q^{12}_{12}-1)\frac{\xi^2}{(4\pi)^2}-\varrho\beta_{Q^{12}_{12}}\frac{\xi^2}{(4\pi)^2}\col\\
\gamma^\varrho_{\tilde{\mathcal{O}}}&=2(\tilde Q-1)\frac{\xi^2}{(4\pi)^2}-\varrho\beta_{\tilde Q}\frac{\xi^2}{(4\pi)^2}
\col
\end{aligned}
\end{equation}
where the first line also follows 
from \cite{Fokken:2014soa} in the limit \eqref{limit}.
The parameter $\varrho$ captures the scheme dependence that starts
at two-loop order and vanishes whenever the respective $\beta$-function
is zero.
The scheme labeled by $\varrho$ is defined by applying minimal subtraction
using $\xi_\varrho=\xi\e^{\frac{\varrho}{2}\varepsilon}$ as coupling constant.
For example, we have $\varrho=0$ in the DR scheme but $\varrho=-\gamma_{\text{E}}+\log4\pi$ in the $\overline{\text{DR}}$ scheme \footnote{In the present case, the DR scheme coincides with the MS scheme.}.

At the fix-point values $Q^{ii}_{ii}=\pm\frac{i}{2}$, $Q^{12}_{12}=\tilde Q=1$, 
the scheme dependence vanishes as expected, and we find that only the first of 
the above anomalous dimensions is non-vanishing but complex and reads 
\begin{equation}
\gamma^\varrho_{\mathcal{O}^{ii}}=\pm2i\frac{\xi^2}{(4\pi)^2}-2\frac{\xi^4}{(4\pi)^4}
\pnt
\end{equation}

\section{Effect on planar multi-point correlation functions}

Not only the two-point functions and hence anomalous dimensions \eqref{gammas}
depend on the double-trace couplings \eqref{doubletracecouplings} in the 
planar limit. Also, planar higher-point functions and hence also OPEs
of operators built from more than two fields are sensitive to these couplings.

\begin{figure}[htbp]
\centering
\centering
\subfigure[]{%
\label{wrapdiag}
\settoheight{\eqoff}{$+$}%
\setlength{\eqoff}{0.5\eqoff}%
\addtolength{\eqoff}{-5.5\unit}%
\raisebox{\eqoff}{%
\begin{pspicture}(-1,-1)(14,18.5)
\ulinsert{1}{16}%
\olvertex{1}{13}%
\dlinsert{1}{10}
\psbezier[doubleline=true](4,16)(5.25,16)(5.25,15)(6.5,15)
\psbezier[doubleline=true](6.5,15)(7.75,15)(7.75,16)(9,16)
\psbezier[doubleline=true](4,13)(5.25,13)(5.25,12)(6.5,12)
\psbezier[doubleline=true](6.5,12)(7.75,12)(7.75,13)(9,13)
\urinsert{12}{16}%
\orvertex{12}{13}%
\drinsert{12}{10}%
\ulinsert{1}{6}%
\olvertex{1}{3}%
\dlinsert{1}{0}
\psbezier[doubleline=true](4,3)(5.25,3)(5.25,4)(6.5,4)
\psbezier[doubleline=true](6.5,4)(7.75,4)(7.75,3)(9,3)
\psbezier[doubleline=true](4,0)(5.25,0)(5.25,1)(6.5,1)
\psbezier[doubleline=true](6.5,1)(7.75,1)(7.75,0)(9,0)
\urinsert{12}{6}%
\orvertex{12}{3}%
\drinsert{12}{0}%
\psset{doubleline=true,linearc=\linearc}
\psline(4,10)(5,10)(5,9)
\psline(5,7)(5,6)(4,6)
\psline(6,8)(7,8)
\psline(8,9)(8,10)(9,10)
\psline(8,7)(8,6)(9,6)
\psbezier(4,8)(-0.5,8)(-0.5,10)(-0.5,13)
\psbezier(-0.5,13)(-0.5,18)(0.5,18)(6.5,18)
\psbezier(6.5,18)(13.5,18)(13.5,18)(13.5,13)
\psbezier(13.5,13)(13.5,10)(13.5,8)(9,8)
\fourvertex{5}{8}{0}
\fourvertex{8}{8}{0}
\end{pspicture}}%
}
\subfigure[]{%
\label{ctdiag}
\settoheight{\eqoff}{$+$}%
\setlength{\eqoff}{0.5\eqoff}%
\addtolength{\eqoff}{-5.5\unit}%
\raisebox{\eqoff}{%
\begin{pspicture}(-1,-1)(14,18.5)
\ulinsert{1}{16}%
\olvertex{1}{13}%
\dlinsert{1}{10}
\psbezier[doubleline=true](4,16)(5.25,16)(5.25,15)(6.5,15)
\psbezier[doubleline=true](6.5,15)(7.75,15)(7.75,16)(9,16)
\psbezier[doubleline=true](4,13)(5.25,13)(5.25,12)(6.5,12)
\psbezier[doubleline=true](6.5,12)(7.75,12)(7.75,13)(9,13)
\urinsert{12}{16}%
\orvertex{12}{13}%
\drinsert{12}{10}%
\ulinsert{1}{6}%
\olvertex{1}{3}%
\dlinsert{1}{0}
\psbezier[doubleline=true](4,3)(5.25,3)(5.25,4)(6.5,4)
\psbezier[doubleline=true](6.5,4)(7.75,4)(7.75,3)(9,3)
\psbezier[doubleline=true](4,0)(5.25,0)(5.25,1)(6.5,1)
\psbezier[doubleline=true](6.5,1)(7.75,1)(7.75,0)(9,0)
\urinsert{12}{6}%
\orvertex{12}{3}%
\drinsert{12}{0}%
\psset{doubleline=true,linearc=\linearc}
\psline(4,10)(5,10)(5,9)
\psline(5,7)(5,6)(4,6)
\psline(8,9)(8,10)(9,10)
\psline(8,7)(8,6)(9,6)
\psbezier(5,9)(5,8.25)(8,8.25)(8,9)
\psbezier(5,7)(5,7.75)(8,7.75)(8,7)
\end{pspicture}}%
}
\caption{The planar four-point function of the operators \eqref{forpointfunction} receives a UV divergent contribution associated with a one-loop vertex correction. The divergence is canceled by the second diagram that contains the counter term of the respective double-trace coupling \eqref{doubletracecouplings}. 
The four-point function is shown in double-line notation and the operators are depicted in red.}
  \label{UVdivergent diagram}
 \end{figure}
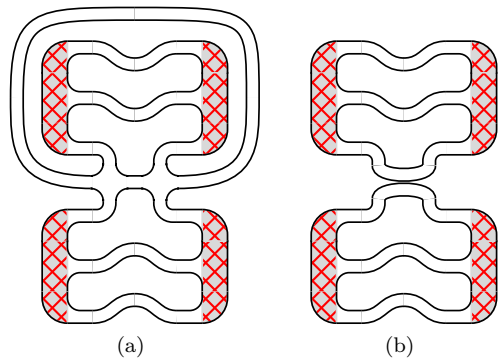

As an example, we consider the four-point correlation function
of the following operators each built from three scalars
\begin{equation}
\label{forpointfunction}
\begin{aligned}
&\tr(\phi_1\phi_2^2)(x_1)\col\qquad
\tr(\phi_1\bar\phi_2^2)(x_2)\col\\
&\tr(\bar\phi_1\phi_2^2)(x_3)\col\qquad
\tr(\bar\phi_1\bar\phi_2^2)(x_4)\pnt
\end{aligned}
\end{equation}
At order $\mathcal{O}(\xi^4)$ and when double-trace couplings are disregarded, 
the diagrams contain two quartic scalar single-trace 
vertices. There is one such diagram that is planar and 
contains a UV divergence 
that cannot be absorbed by renormalizing the operators \eqref{forpointfunction}.
It contains a loop formed by the two direct connections of these
two vertices and is shown in figure \ref{wrapdiag}.
Its UV divergence is associated with the one-loop correction of a quartic 
scalar double-trace vertex and has to be canceled by a respective counter term diagram 
that also contributes in the planar limit. The latter diagram is shown in figure \ref{ctdiag} 
 and contains the respective counter 
term coupling from the double-trace Lagrangian
\eqref{doubletracecouplings}  \footnote{For non-vanishing tree-level double-trace couplings, also a third diagram with two double-trace couplings exists. It does not affect the argument.}. 
The result is therefore sensitive
to the $\beta$-function of that coupling. This is another 
indication that one must not neglect the double-trace couplings 
-- not even in the planar limit. 

Diagrams similar to those in figure \ref{UVdivergent diagram} exist for 
all correlation functions of composite single-trace operators 
in which the total charge of a subset of these operators matches the charge 
of a single-trace factor in a double-trace coupling. 
Likewise, a composite operator 
$\mathcal{O}^{ii}$, $\mathcal{O}^{12}_{12}$, $\tilde{\mathcal{O}}$ or a complex
conjugate thereof can occur in the
OPE of two single-trace operators whose total charge matches its charge.
Via the anomalous dimension for that operator taken from \eqref{gammas}, 
such an OPE is then sensitive to the respective 
$\beta$-function.
The planar correlation functions of the field theory are hence widely affected by the
double-trace couplings \eqref{doubletracecouplings}.

\section{Conclusions}

In this letter, we have explicitly shown that the model proposed in 
\cite{Gurdogan:2015csr} is neither complete nor conformal -- not even in the planar 
limit. 
We have shown how to complete the model in the planar limit by including  
the required double-trace couplings and their counter terms.
Admitting complex coupling constants, we could find conformal fix-points
for all induced couplings at one-loop order. 

It would be very interesting to determine whether the fix-points 
persist at higher loop orders and at finite $N$, i.e.\ when terms beyond the 
planar limit that are subleading in the large-$N$ expansion are taken 
into account.

Finally, the model opens the possibility for very interesting studies 
concerning integrability in a simplified setup which 
are of high relevance for integrability in $\mathcal{N}=4$ SYM theory
and in its deformations. 
For instance, the very interesting question whether integrability is connected to 
conformal invariance can be investigated by studying the spectrum of 
the operators $\mathcal{O}^{ii}$ and $\mathcal{O}^{12}_{12}$ or 
$\tilde{\mathcal{O}}$ at the conformal 
fix-points and away from them. In particular, it should be analyzed 
whether the different behaviors (non-vanishing and vanishing) of their 
anomalous dimensions \eqref{gammas} at the fix-points can be reproduced in 
the integrability-based approach. 
This analysis should also be extended to both classes of
multi-point functions: those that are sensitive and those that are not 
sensitive to the breakdown of conformal invariance. 

\section{Acknowledgment}

We thank M.\ Staudacher and K.\ Zarembo for reading the manuscript and 
for very useful remarks.
This research is supported in part by SFB 647 \emph{``Raum-Zeit-Materie. Analytische und Geometrische Strukturen''} and in part by DFF-FNU  through
grant number DFF-4002-00037.

\bibliography{refs}

\end{document}